\documentstyle[twoside,fleqn,espcrc2,epsfig]{article}

\newcommand{\AmS}{{\protect\the\textfont2
  A\kern-.1667em\lower.5ex\hbox{M}\kern-.125emS}}

\title{M Theory Fivebrane and SQCD\thanks{Talk presented at Strings
'97 in Amsterdam, Holland (June 16 - 21, 1997).}}

\author{Hirosi Ooguri
\\ ~~\\
366 Le\thinspace Conte Hall, Department of
Physics \\ University of California at Berkeley, Berkeley, CA 94720,
        USA \\ and \\
Theory Group, Mail Stop 50A-5101, Physics Division \\
Lawrence Berkeley National Laboratory, Berkeley, CA 94720, USA.}
       
\begin{document}


\maketitle

\section{Introduction}
A low energy effective theory of parallel D(irichlet) branes is
a gauge theory with sixteen supercharges, but one can consider a web
of brane to realize situations with reduced number of supersymmetry
\cite{HW}. In this talk, I will discuss four-dimensional theories with
$N=1$ and $2$ supersymmetry (i.e. four and eight supercharges). 
In the case of theories with $N=2$ supersymmetry, the exact 
description of the Coulomb branch is given by reinterpreting
the web of branes as a configuration of a single fivebrane
in the IIA theory \cite{Geom,W2}. Recently we studied
the case with $N=1$ supersymmetry, and found that description
in term of the fivebrane in M Theory captures strong coupling
dynamics of the $N=1$ gauge theory in four dimensions \cite{HOO}.
In particular, we found that the configuration
of the fivebrane geometrically encodes information on the
Affleck-Dine-Seiberg superpotential and the structure of the
quantum moduli space of vacua. Simultaneously to our work, the 
case without matter field was studied in \cite{W1}. A related
work also appeared in \cite{IGT}.   

\section{Geometric Engineering of $N=2$ Gauge Theory}

\subsection{Web of Branes}
Let me first describe how to construct $N=2$ gauge theory from
a web of branes in the type IIA theory. We first separate the ten
dimensions into six and four. The $\bf{R}^{3,1}$ part of the spacetime
is parallel to the worldvolume of all the branes, and therefore it
is where the gauge theory is realized. We then tie together the
branes in the six dimensions as shown in figure 1 so that 
the desired field content is realized. 
\begin{figure}[htb]
\begin{center}
\epsfxsize=3in\leavevmode\epsfbox{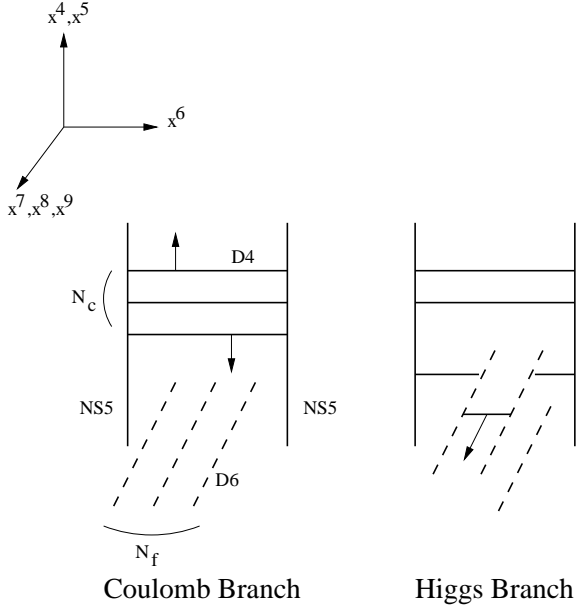}
\end{center}
\caption{Coulomb and Higgs branches in the IIA picture}
\label{fig1}
\end{figure}

To realize the theory with $SU(N_c)$ gauge group, we consider $N_c$
D4 branes suspended between two parallel NS5 branes. The open string
attached to the D4 branes gives rise to the gauge boson of
$SU(N_c)$. We then introduce $N_f$ D6 branes perpendicular to the D4 and
NS5 branes, and the open string going between the D4 and D6 branes
give $N_f$ matter fields in the fundamental representations, which I will 
call quarks. The moduli space of $N=2$ vacua consists of the
Coulomb and Higgs branches (and also their mixed branches). The
Coulomb branch is parametrized by the vacuum expectation value (vev) 
of the adjoint scalar field in the $N=2$ vector multiplet while the
Higgs branch is parametrized by the squarks. In the IIA picture, they
are described by locations of the D4 branes. The motion of the
D4 branes along the NS5 brane parametrizes the Coulomb branch.
To go from the Coulomb to the Higgs branch, one has to move 
the D4 brane to the D6 brane, break the D4 brane on the D6 brane,
and then move a segment of the D4 brane along the D6 brane. Note that
in order to do this, we have to have at least two D6 branes and they
have to be located at the same point in the $x^4, x^5$ direction (see
figure 1). This corresponds to the field theory fact that the Higgs
branch requires at least two massless matter fields.

\subsection{Lifting to M Theory}

We can lift this configuration to M Theory \cite{W2}. As it will become
clear later, this will enable us to incorporate nonperturbative
gauge theory dynamics in the brane picture. The D4 and NS5 branes 
of the IIA theory come from the fivebrane of M theory. When
the fivebrane is wrapped on the $S^1$ of the eleventh dimension,
it becomes the D4 brane. When it is not wrapped, it gives
the NS5 brane. On the other hand, D6 brane localized at points
in the $(x^4, x^5, x^6)$ plane is interpreted as the
Kaluza-Klein monopole and is given by the Taub-NUT metric
\begin{eqnarray}
ds^2 = V(x) dx^i dx^i + V(x)^{-1} (dx^{10} + A_i(x) dx^i)^2 
\nonumber \\ 
V(x) = 1 + \sum_n \frac{R}{|x-x_n|},~~
   \nabla V = \nabla \times A, 
\label{Taub}
\end{eqnarray}
where the D6 branes are located at $x^i=x_n^i$ ($i=4,5,6$) and
$x^{10}$ is the eleventh dimensional coordinate with the periodicity
$x^{10} \simeq x^{10} + 2 \pi R$ \cite{town}. 

Let us now discuss the configuration of the fivebrane. 
As in the IIA picture, we have to keep four dimensions of 
the six-dimensional worldvolume of the fivebrane to be flat
$\bf{R}^{3,1}$. In order to preserve the $N=2$ supersymmetry
in four dimensions, the remaining two dimensional part has
to be embedded holomorphically in the Taub-NUT geometry
and is localized in the $x^7, x^8, x^9$ directions. (If
the fivebrane is also extended in those directions, it
will have generically smaller supersymmetry, i.e. $N=1$
in four dimensions. I will discuss more about it later.)
So let us introduce complex coordinates,
\begin{equation}
  v = x^4 + ix^5,~~ t = {\rm exp}(-\frac{x^6+ix^{10}}{R}).
\end{equation}
Now we can transform the IIA picture into the M theory picture.
The fivebrane must have two asymptotic regions corresponding to
the two infinite NS 5-branes. In the middle, there are
$N_c$ branches of the brane corresponding to the D4 branes.
We also need $N_f$ D6 branes, which are now represented by
the Taub-NUT geometry (\ref{Taub}). It was shown in \cite{W2} that
such a fivebrane configuration is unique and is the same
as the Seiberg-Witten curve \cite{sw,sw2},
\begin{equation}
   t^2 - B(v) t + \Lambda_{N=2}^{2N_c-N_f} v^{N_f} = 0,
\label{s-w}
\end{equation}
where $B(v) = v^{N_c} + \cdots$ is a polynomial of degree
$N_c$ in $v$ and depends on the Coulomb branch moduli of the
theory. 

\subsection{Why does this work?}

Since the eleven-dimensional supergravity is supposed to describe
the low energy dynamics of the strong coupling limit of the type IIA
theory, it is reasonable to expect that the reinterpriting the web
of branes in the IIA theory as a configuration of the single fivebrane
captures the nonperturbative physics of the gauge theory, and in fact
we saw in the above that it is the case for the $N=2$ gauge theory. 
This statement needs some clarification and it is useful to 
understand exactly how it works.

In the IIA picture, the gauge coupling constant $\lambda$ is related
to the string coupling constant $g$ as
\begin{equation}
   \lambda^2 = \frac{g l_s}{L}
\label{coupling}
\end{equation}
where $l_s$ is the string scale set by the string tension and
$L$ is the distance between the NS 5-branes. The $L$ factor appears
in the formula since the gauge theory lives on the D4 brane
worldvolume, which is ${\bf R}^{3,1}$ times the line segment in the $x^6$
direction. If $1/L$ is larger than the typical gauge theory scale,
we can ignore the Kaluza-Klein type excitation in the $x^6$ direction
and the low energy theory is on ${\bf R}^{3,1}$. The volume
$L$ then renormalizes the gauge coupling constant as in
(\ref{coupling}). Since the radius of the eleventh dimensional $S^1$
is $R = g l_s$, we can also express $\lambda$ as
\begin{equation}
   \lambda^2 = \frac{R}{L}. 
\end{equation}

To relate the low energy theory on the D4 branes to the standard
gauge theory in four dimensions, we have to make some
limit.

\noindent
(1) We need to take the string coupling constant $g$ to be small
in order the theory on the brane to decouple from physics
of the bulk ten or eleven dimensions. In particular, we should
be able to ignore gravitational effects in the bulk in order to
concentrate on physics on the brane. This is possible if we send
$g \rightarrow 0$, or equivalently $R \ll l_s$.

\noindent
(2) On the other hand, we want $\lambda$ to be finite in order to be
able to observe interesting strong coupling physics of the gauge
theory.

\noindent
(3) These conditions require $L \ll l_s$. In fact this is
also required in order to be able to ignore the Kaluza-Klein type mode
on the brane, as I mentioned in the above. 

So we need to take the limit, $R, L \ll l_s$, while keeping
$\lambda^2 = R/L$ finite. 

On the other hand, the eleven-dimensional supergravity gives
a good description in the limit $l_{p} \ll R, L$ where
$l_p = g^{1/3} l_s$ is the eleven-dimensional Planck length,
and this means $l_s \ll R, L$, which is complete opposite of the above
limit.

Nevertheless the M theory fivebrane gives the correct description
of the $N=2$ Coulomb branch, as we saw in section 2.2. The reason for this
is the same as the one used to derive exact results on  
Calabi-Yau compactification \cite{strominger,Geom2,Geom}, 
i.e. the decoupling of the vector and hypermultiplet fields. 
In the present case, parameters characterizing the size of the 
brane are in hypermultiplets of the four-dimensional theory 
while its shape is determined by vector multiplet fields. 
Since the Coulomb branch of the four-dimensional gauge theory
is parametrized by the vector multiplet fields, it should depend
on $R$ and $L$ only through the ratio $L/R$. This means
that, although we have to take both $R$ and $L$ to be small 
in order to describe the gauge theory in four dimensions, since
the theory depends only on their ratio, 
the same result is obtained by taking them to be large as far as we
keep the ratio fixed. This is why we can trust
the eleven-dimensional supergravity to give the correct low
energy description. 

\subsection{Relation to Calabi-Yau Compactification} 

It has been known earlier in the context of the Calabi-Yau
compactification \cite{Geom2} that the exact solution of the $N=2$
gauge theory arises from the classical geometry of string
compactification if one takes the field theory limit in such a
way that gravity is turned off while the gauge theory scale is held
fixed (this involves taking the singular limit of the Calabi-Yau
manifold).  It was then found in \cite{Geom} that such a  Calabi-Yau manifold 
in the IIB case is T-dual to a configuration of the NS 5-brane in the IIA
theory, which is exactly given by (\ref{s-w}). The T-duality of a
singular Calabi-Yau manifold and a NS 5-brane configuration was
pointed out in \cite{OV} and was further clarified recently in
\cite{GHM}. 

In the following, we will use the M theory fivebrane to describe the
gauge theory with $N=1$ supersymmetry. We will see the strong coupling
physics of gauge theory is geometrically engineered in the $N=1$ case.
It is reasonable to expect that such a fivebrane configuration is
related to the F-theory compactification \cite{vafaftheory,1geom} 
on a Calabi-Yau four-fold
\footnote{I would like to thank Cumrun Vafa for discussion on
this.}. It would be very interesting to reinterpret the results of
our work from the point of view of the F-theory.

\section{Rotation to N=1}

\subsection{IIA Picture}

Now we break the $N=2$ supersymmetry to $N=1$ by adding a mass $\mu$
for the adjoint chiral multiplet in the $N=2$ vector multiplet. 
\begin{figure}[htb]
\begin{center}
\epsfxsize=3in\leavevmode\epsfbox{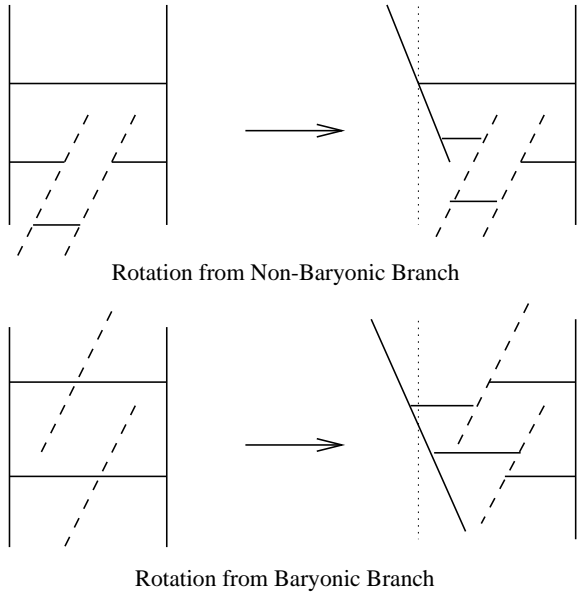}
\end{center}
\caption{Rotation of the left NS 5-brane in the IIA picture}
\label{fig2}
\end{figure}
Since the adjoint scalar becomes massive, most of the Coulomb branch
is lifted except at points on the roots of the Higgs
branches \cite{aps} and some other exceptional points in the
Coulomb branch \cite{HOO}. Let us first view this
procedure in the IIA picture. It was pointed out by Barbon
\cite{barbon} that adding the adjoint mass is the same as changing
the relative orientation of the two NS5 branes. Let us see how
it looks like in the examples of $SU(2)$ with 2 flavors. 
If the two D4 branes are at generic positions,
we cannot rotate the NS5 branes without breaking the supersymmetry
completely. In this case, there are two situations in which
the rotation is possible, as shown in figure 2. One is at the
non-baryonic branch root. Note that the location of one of
the D4 brane in figure 2 is fixed to be at the axis of the rotation.
This means that we have to start at particular points on the
non-baryonic root. 
In general, these are points where there are maximum number of
mutually local monopoles. Another possibility is at the baryonic
branch root, where all the D4 branes are fixed on the D6
branes from the begining. We can then break the D4 branes
on the D6 branes, as shown in figure 2, and rotate the configuration.
In the limit when the two NS5 branes make the right angle,
we recover the configuration introduced in \cite{egk} to
study $N=1$ SQCD.

\subsection{M Theory Picture}

Now we come to one of the main points of this talk that is to
reinterpret this in the M theory picture. Let me first describe it
using the same example of $SU(2)$ gauge theory with 2 flavors.

In the M theory description, the fivebrane at the rotable point on
the non-baryonic branch root is given by the equation
\begin{equation}
    t^2 - v^2 t + \Lambda^2 v^2 = 0.
\label{rotatablecurve}
\end{equation} 
The corresponding curve is drawn in figure 3, in the dotted line in the 
$v-t$ plane. 
\begin{figure}[htb]
\begin{center}
\epsfxsize=3in\leavevmode\epsfbox{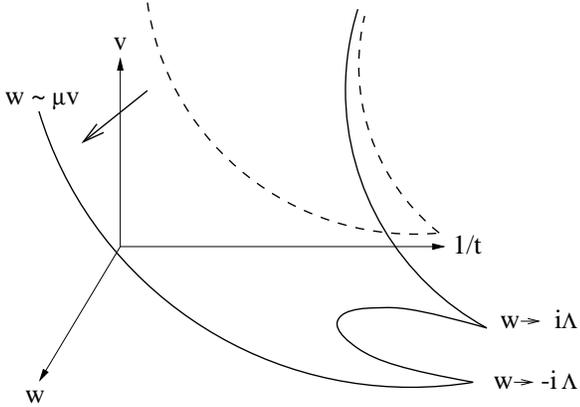}
\end{center}
\caption{Brane Rotation from Non-Barynonic Branch}
\label{fig3}
\end{figure}
The two branches extending to infinities correspond
to the two NS5 branes. Therefore the brane rotation that I
described in the previous subsection should correspond to 
pulling one of the branches out of the $v-t$ plane and toward the 
$w = x^6 + i x^7$ direction. We have found a unique way to do this,
and it is described by the following equations in the $v-t-w$ plane.
\begin{eqnarray}
wv = \mu^{-1} (w^2 + \mu^2 \Lambda^2) \nonumber \\
t = \mu^{-2}(w^2 + \mu^2 \Lambda^2) . 
\end{eqnarray}
One can easily check that this solves (\ref{rotatablecurve}) for any
value of $w$. 
The corresponding curve is drawn in the solid line in figure 3. 
We note that the left branch is now extended into $w = \mu v$
direction, where $\mu$ is the adjoint mass while the right
branch asymptotically approaches the original curve. We also note
that there are two additional asymptotic regions, with different
values of $w$. This will become important later when
we compare this brane configuration with field theory results.    

At the baryonic branch root, the equation factorized into
two components,
\begin{equation}
 (t-v^2)(t - \Lambda^2) = 0,
\end{equation}
and this means that the curve factorizes into two components
as shown in the dotted line in figure 4. 
\begin{figure}[htb]
\begin{center}
\epsfxsize=3in\leavevmode\epsfbox{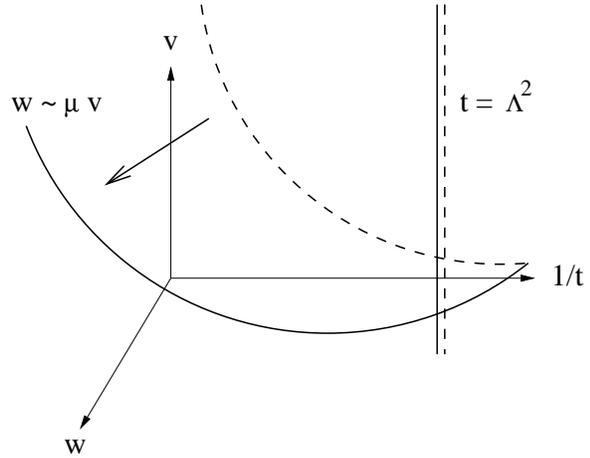}
\end{center}
\caption{Brane Rotation from Barynonic Branch}
\label{fig4}
\end{figure}
We can then easily
rotate the curve as
\begin{eqnarray}
~(1)~~ t = v^2, ~~w = \mu v \nonumber \\
(2)~~ t = \Lambda^2, ~~ w=0, \nonumber
\end{eqnarray}
and the corresponding curve is depicted in the solid line
in figure 4. 

Let us generalize this to $SU(N_c)$ gauge theory with
$N_f$ quarks, for arbitrary $N_c$ and $N_f$. First I want to
point out that the brane is rotatable if and only if the configuration
is birationally equivalent to $\bf{CP}^1$.
To show this, it is convenient to 
regard $w$ of the rotated curve as a function along the original
curve. This is always possible for small values of $\mu$ and it
can be shown to be true for any $\mu$ by using the R symmetry
(see \cite{HOO}). The brane configuration after the rotation
then can be regarded as a graph of $w$. We note that $w$ is
infinite only at one point, corresponding to the left NS5 brane
extending to the $w=\mu v$ direction. Since $1/v$ is a good coordinate
on that branch, it should be regarded as a simple pole. In order 
to have a holomorphic function $w$ with a simple pole only at
one point, the original curve must have been equivalent to
$\bf{CP}^1$. I used the word ``birationally'' in the above
since the original curve may have two or more points coinciding.
Such points are separated after the rotation since any
two points must have different values of $w$. That the original
curve has to be genus $0$ is consistent with what we expect
from the field theory analysis \cite{sw,aps}. The Seiberg-Witten
curve of the $N=2$ theory that survives the adjoint mass
perturbation has exactly this property. 

This observation also makes it possible to find the brane configuration
explicitly, for any $N_c$ and $N_f$, after the rotation. This
is because $w$ as a function with a simple pole, is unique
up to $SL(2,\bf{Z})$ on $\bf{CP}^1$, and $SL(2,\bf{Z})$
is fixed by the asymptotic conditions on the curve. 

If one starts with the $r$-th non-baryonic branch root
$r=1,...,[N_f/2]$, the rotated curve is given explicitly 
by\footnote{For $N_f \geq N_c$, $r=N_f-N_c$ has to be excluded from 
rotatable curves \cite{HOO}.}
\begin{eqnarray}
vw = \mu^{-1} (w-w_+)(w-w_-) ~~~~~~~~~~~~~ \nonumber \\
 t = \mu^{-N_c} w^{N_c-N_f}(w-w_+)^r (w-w_-)^{N_f-r} ,
\end{eqnarray}
where
\begin{equation}
  w_\pm = \left[ (-1)^{N_c+r_\pm} \left( \frac{N_c-r_-}{N_c-r_+} 
   \right)^{N_c-r_\pm} \right]^{\frac{1}{2N_c-N_f}} \mu
   \Lambda,
\label{branches}
\end{equation}
and $r_+=r$, $r_-=N+f-r$. It turned out there is another rotatable
curve which is obtained by setting $r=0$ in the above, and there
is a corresponding point in the Coulomb branch which survives the
perturbation by the adjoint mass term. This point 
is not attached to any Higgs branch and therefore is not in the list
of \cite{aps} where they studied points on the Higgs branch roots.

As before, the left branch of the curve extends to $w=\mu v$
direction,
while the right branch approaches the original curve. There are
additional
two asymptotic regions, with fixed values of $w = w_+$ and $w_-$.
These correspond to locations of D4 branes in the IIA picture.

The rotation from the baryonic branch can be easily carried out
since the curve factorizes there. The result of the rotation is
\begin{eqnarray}
~(1)~~ t = v^{N_c}, ~w=\mu v ~~~~~~~~~~~~~~~~ \nonumber \\
(2)~~ t = \Lambda^{2N_c-N_f} v^{N_f-N_c}, ~w=0 . 
\end{eqnarray}

\section{Comparison with Field Theory Results}

\subsection{Affleck-Dine-Seiberg Superpotential}

Now I would like to compare these with field theory results. 
When $N_c > N_f$, it is known that the $N=1$ theory dynamically
generates the Affleck-Dine-Seiberg superpotential. At $N_c-1 = N_f$,
it is generated by the instanton effects and for $N_c-1 > N_f$ by some
strong coupling effects.

Let us start with the $N=2$ theory and add the adjoint mass $\mu$.
Before we turn on the adjoint mass, the superpotential of the $N=2$
theory is ${\rm tr} Q \Phi \tilde{Q}$ where $\Phi_{ij}$ is the
adjoint field and $Q_{ia}$, $\tilde{Q}_{ia}$ are quarks
($i,j=1,...,N_f$, $a,b = 1,...,N_c$). If $\mu$ is much larger
that the scale $\Lambda$ of the $N=2$ theory, we can integrate out
$\Phi$ and generate the superpotential term
\begin{equation}
   \frac{1}{2\mu} \left[ {\rm tr} M^2 - \frac{1}{N_c}
 ({\rm tr}M)^2 \right],
\end{equation}
where $M_{ij} = (\tilde{Q}Q)_{ij}$ is the meson made out of the
quarks. In the $\mu \rightarrow \infty$ limit, we obtain the
$N=1$ SQCD where we know that the Affleck-Dine-Seiberg potential
is generated. For large $\mu$, therefore, we expect that
the effective superpotential is given by
\begin{eqnarray}
W_{eff} &=& (N_c -N_f) \left( \frac{\Lambda_{N=1}^{3N_c-N_f}}{{\rm det}
M} \right)^{\frac{1}{N_c-N_f}} +\nonumber \\
&&  +\frac{1}{2\mu} \left( {\rm tr} M^2 -
\frac{1}{N_c} ({\rm tr}M)^2 \right), 
\end{eqnarray}
where $\Lambda_{N=1}$ is the scale of the $N=1$ theory set by
the renormalization group matching condition
\begin{equation}
  \Lambda_{N=1}^{3N_c-N_f} = \mu^{N_c} \Lambda^{2N_c - N_f}.
\end{equation}
In fact, by using the standard holomorphy argument \cite{is},  
one can prove that the superpotential $W_{eff}$ is exact
for any $\mu$. 

In the $\mu \rightarrow \infty$ limit, the $N=1$ theory has no
vacua since $W_{eff}$ has no minima (the vacuum runs away). 
The $\frac{1}{\mu}$ term in $W_{eff}$ stabilizes the vacua
for finite $\mu$, so we can analyse its structure. By extrematizing
$W_{eff}$, we can determine vev
 of the meson field $M_{ij}$. 
One can show \cite{HOO} that $M$ solving
\begin{equation}
  \frac{\partial W_{eff}}{\partial M_{ij}} = 0 
\end{equation}
is diagonalizable by a unitary matrix, and have
two different eigenvalues, $m_+$ and $m_-$.
Let $r$ be the number of $m_+$ ($r=0,...,[N_f/2]$).
The eigenvalues are then give by
\begin{equation}
 m_\pm =  \left[ (-1)^{r_\pm} \left( \frac{N_c-r_-}{N_c-r_+}
   \right)^{N_c-r_\pm} \right]^{\frac{1}{2N_c-N_f}} \mu
   \Lambda.
\end{equation}
As one can see by comparing with (\ref{branches}), 
the eigenvalues $m_\pm$ of the meson field $M$ are equal
to the asymptotic locations $w_\pm$ of the branches of the
rotated fivebrane,
upto a trivial overall phase factor independent of $r$. 

This means that the information on the Affleck-Dine-Seiberg
potential is {\it geometrically encoded} in the fivebrane
configuration. In the IIA picture, the vev of the meson field
is specified by the locations of the D4 branes along the D6 branes.
In the M theory picture, these are exactly the locations of
the fivebrane branches characterized by $w_\pm$. Not only their
values agree with the field theory computation of $M$ in the
above, but their multiplicities match (the fivebrane is wrapping
the two asymptotic directions $r$ and $(N_f-r)$ times respectively,
and these numbers
agree with the multiplicities of the eigenvalues of M). 
 
So far we have discussed the case with massless quarks.
We can add massed to the quarks and perform the same analysis. We
found  a complete agreement between the field theory analysis
and the brane configuration. 

\subsection{Quamtum Moduli Space}

For $N_c \leq N_f$, there are stable vacua in 
the $\mu \rightarrow \infty$ limit and their moduli space is of
interest \cite{is}. Classically the moduli space is parametrized by
gauge invariant polynomials of the squarks obeying the constraints. 
When $N_c = N_f$, the moduli space is deformed by the quamtum
corrections.  On the other hand, for $N_c < N_f$,  the quamtum moduli
space at $\mu \rightarrow \infty$ is the same as the classical
one. For finite $\mu$, however interesting quamtum effects appear even
for $N_c < N_f$. In the paper \cite{HOO}, we have shown that one can read
off the vevs of the gauge invariant polynomials from the configuration of the
fivebrane and found complete agreement with the field theory
predictions. For more detail, I would like to refer the reader to the
original paper. 

\section{Conclusion}

We have found that the classical eleven-dimensional supergravity can
be used to study strong coupling dynamics of $N=1$ SQCD in four
dimensions. Nonperturbative corrections to the superpotential and the
moduli space structure are geometrically encoded in the configuration
of the fivebrane. First of all, this result provides a very strong
evidence for the conjecture that the strong coupling limit of the IIA
string theory is the eleven-dimensional theory. 
This geometrical description may also give us new
insights into nonperturbative effects in the gauge theory in four dimensions. 

I should point out that the dynamical information on
the gauge theory that we have extracted from the fivebrane is
holomorphic in nature, i.e. the holomorphic superpotential and the
holomorphic property of the moduli space. For these, an argument
similar to that given in section 2.3 guarrantees that the classical
supergravity computation captures the corect gauge theory dynamics in four
dimensions. To study non-holomorphic aspects of the gauge theory, one
needs to control possible corrections to the classical fivebrane
description. There are potentially two sources of corrections.  The
classial supergravity approximation may fail when the curvature induced
by the fivebrane becomes strong compared to the Planck scale. Other
corrections may arise if Kaluza-Klein modes on the curve $\Sigma$
or bulk degrees of freedom do not decouple from the four-dimensional
gauge theory. It would be important to identify the region of validity
of the supergravity description. Related issues were discussed in
Maldacena's talk at this conference \cite{malda}. 

\bigskip

\noindent
{\bf Acknowledgements}

\smallskip

It is my pleasure to thank the organizers of Strings `97 
for their hospitality and for 
giving me the opportunity to present this work at the beautifully
organized conference. I would like to thank
Kentaro Hori and Yaron Oz for the collaboration on this work.  I would
also like to thank Cumrun Vafa for collaborations on related works and
for useful discussions. 

This research is supported in part by NSF grant PHY-95-14797 and DOE
grant DE-AC03-76SF00098.

\end{document}